\documentclass[english]{sbrt}
\usepackage[english]{babel}
\usepackage[utf8]{inputenc}
\usepackage{graphicx}
\usepackage[svgnames]{xcolor}
\usepackage{amsmath,amsfonts,amssymb}
\usepackage[ruled,vlined]{algorithm2e}
\usepackage{algpseudocode}
\usepackage[nolist,nohyperlinks]{acronym} 
\usepackage{subfig}

 \newcommand{\ten}[1]{\boldsymbol{\mathcal #1}}
 \newcommand{\ma}[1]{\boldsymbol{#1}}
 
\begin{document}
\begin{acronym}
    \acro{MIS}{Movable intelligent surface}
	\acro{IRS}{Intelligent reflecting surface}
	\acro{RIS}{Reconfigurable intelligent surface}
    \acro{SIM}{Stacked intelligent metasurface}
    \acro{FAS}{Fluid antenna system}
    \acro{RIs}{reconfigurable intelligent surface}
	\acro{irs}{intelligent reflecting surface}
	\acro{PARAFAC}{parallel factor}
	\acro{TALS}{trilinear alternating least squares}
	\acro{BALS}{bilinear alternating least squares}
    \acro{ALS}{alternating least squares}
	\acro{DF}{decode-and-forward}
	\acro{AF}{amplify-and-forward}
	\acro{CE}{channel estimation}
	\acro{RF}{radio-frequency}
	\acro{THz}{Terahertz communication}
	\acro{EVD}{eigenvalue decomposition}
	\acro{CRB}{Cramér-Rao lower bound}
	\acro{CSI}{channel state information}
	\acro{BS}{base station}
	\acro{MIMO}{multiple-input multiple-output}
	\acro{NMSE}{normalized mean square error}
	\acro{2G}{Second Generation}
	\acro{3G}{3$^\text{rd}$~Generation}
	\acro{3GPP}{3$^\text{rd}$~Generation Partnership Project}
	\acro{4G}{4$^\text{th}$~Generation}
	\acro{5G}{5$^\text{th}$~Generation}
	\acro{6G}{6$^\text{th}$~generation}
	\acro{E-TALS}{\textit{enhanced} TALS}
	\acro{UT}{user terminal}
	\acro{UTs}{users terminal}
	\acro{LS}{least squares}
	\acro{KRF}{Khatri-Rao factorization}
	\acro{KF}{Kronecker factorization}
	\acro{MU-MIMO}{multi-user multiple-input multiple-output}
	\acro{MU-MISO}{multi-user multiple-input single-output}
	\acro{MU}{multi-user}
	\acro{SER}{symbol error rate}
	\acro{SNR}{signal-to-noise ratio}
	\acro{SVD}{singular value decomposition}
    \acro{DFT}{Discrete Fourier Transform}
\end{acronym}
\title{Channel Estimation for Movable Intelligent Surface}

\author{Daniel C. Alcantara, Josué V. de Araújo, Gilderlan T. de Araújo and André L. F. de Almeida
\thanks{Daniel C. Alcantara, Josué V. de Araújo and André L. F. de Almeida are with the Teleinformatics Department, Federal University of Ceará, Fortaleza-CE, e-mails: danielchaves@alu.ufc.br, josue.vas@alu.ufc.br; andre@gtel.ufc.br.}
\thanks{Gilderlan T. de Araújo is with the Federal Institute of Ceará, e-mail: gilerlan.tavares@ifce.edu.br.}
\thanks{This work is partially supported by the National Institute of Science and Technology (INCT-Signals) sponsored by Brazil’s National Council for Scientific and Technological Development (CNPq) (Proc. 406517/2022-3), and FUNCAP (Proc. INCT-25255-82587.32.41/64). The research of André L. F. de Almeida is partially supported by CNPq (Proc. 303356/2025-1). The research of Gilderlan T. de Araújo is supported by CNPq (Proc. 151870/2025-0).}%
}

\maketitle


\renewcommand\baselinestretch{.93}

\begin{abstract}
This paper proposes a tensor-based channel estimation framework for an uplink MIMO system assisted by a movable intelligent surface. The considered architecture combines a fixed transmissive metasurface with a smaller movable layer, whose discrete positions create an additional structured training dimension. By jointly exploiting fixed-layer phase patterns and movable-layer positions, the received pilots are modeled as a fourth-order PARAFAC tensor. A trilinear alternating least-squares receiver is then derived to estimate the individual channels and the position-dependent response. Importantly, the proposed method does not require prior knowledge of the movable-layer phase response at the receiver, since this unknown factor is estimated from the tensor structure of the received signal. Simulation results show that increasing the training length improves the NMSE of the estimated factors and the reconstructed cascaded channel.
\end{abstract}

\renewcommand\baselinestretch{.83}

\section{Introduction}
Future wireless networks are expected to provide high data rates, extended coverage, and energy-efficient operation under increasingly complex propagation conditions. In this context, \acp{RIS} have emerged as a key technology for creating programmable radio environments by controlling the phase, amplitude, and polarization of impinging electromagnetic waves through nearly passive metasurface elements \cite{Di_Renzo_2020}, \cite{RISRuiZhang}. Conventional single-layer \acp{RIS} can improve coverage and spectral efficiency with low power consumption, but its performance depends on accurate passive beamforming and reliable \ac{CSI}.

Several metasurface architectures have been proposed to increase the flexibility of \ac{RIS}-aided systems. Active \acp{RIS} introduces amplification to compensate for the double path-loss effect, at the cost of additional power consumption and hardware complexity \cite{Wang_2025}. \acp{SIM} stacks multiple programmable metasurface layers to perform wave-domain signal processing and enable more advanced MIMO precoding/combining operations \cite{SIM_MIMO_2024}, \cite{SIM_CE_2024}. In parallel, \acp{FAS} and movable-antenna concepts exploit position reconfiguration to access spatial channel variations and enhance diversity without relying only on conventional fixed-position arrays \cite{FAS_Tutorial_2024}. These developments reveal a common trend: future metasurface-aided systems are moving from purely static reflecting panels toward architectures that introduce new physical degrees of freedom.

Among these emerging architectures, the \ac{MIS} proposed in \cite{paperMIS} offers an attractive compromise between beamforming flexibility and hardware simplicity. The \ac{MIS} consists of two transmissive metasurfaces: a larger fixed layer, denoted by MS1, and a smaller movable layer, denoted by MS2. By mechanically shifting MS2 relative to MS1, different subsets of elements are spatially aligned, producing distinct effective phase profiles while keeping the individual phase shifts constant. Therefore, dynamic beam pattern synthesis is achieved through differential position shifting rather than through fast element-wise electronic tuning.

Despite this appealing architecture, channel estimation remains a major bottleneck for \ac{MIS}-assisted \ac{MIMO} systems. As in conventional \ac{RIS}-aided systems, the passive nature of the metasurface prevents direct baseband observation at the surface, making the individual \ac{UT}-\ac{MIS} and \ac{MIS}-\ac{BS} channels difficult to acquire. Moreover, the movable layer introduces an additional position-dependent response, which must be estimated or calibrated jointly with the wireless channels. Existing works on \ac{RIS} channel estimation include training-based, semi-blind, and tensor-based strategies \cite{Zhang_2022}, \cite{Lee_2022}, \cite{Gil_JTSP}, \cite{Paulo_2023}. By retaining the natural coupling across dimensions such as space, time, and coding, tensor models enable parameter estimation with reduced training overhead. Early studies established classical tensor decompositions as effective tools for blind and semi-blind estimation in conventional MIMO systems and space-time-frequency coding schemes \cite{Almeida2007,confac,FAVIER_2012,Favier_Almeida2013,Favier_TSP_2014}. In the context of RIS-assisted networks, tensor-based techniques have been successfully applied to decouple cascaded channels \cite{AraujoSAM2020,Gil_JTSP,Wei2021,ardah2021trice}. For example, PARAFAC-based models have been used to estimate cascaded and individual channels in \ac{RIS}-assisted MIMO systems \cite{Gil_JTSP}, while two-timescale strategies exploit different fading rates to reduce pilot overhead \cite{twotimeScale}.

However, these solutions do not directly address the additional mechanical degree of freedom of \ac{MIS}-aided MIMO systems. In particular, the position-dependent phase response generated by MS2 introduces an additional structured dimension that can be exploited for channel estimation but also imposes new identifiability and training-design constraints. This setting is especially challenging from a practical standpoint because the phase response induced by the movable layer may be unknown to the receiver. In the proposed approach, this unknown movable-layer phase matrix is not assumed to be calibrated a priori; instead, it is treated as an additional tensor factor and estimated jointly with the wireless channels. To the best of our knowledge, the channel estimation problem for \ac{MIS}-assisted MIMO architectures has not yet been addressed in the literature. 
The contributions of this work are summarized as follows:
\begin{itemize}
    \item We propose a two-dimensional training protocol that combines fixed-layer phase-shift patterns and movable-layer positions, thereby modeling the received signal as a fourth-order PARAFAC tensor.
    
    \item We derive a TALS-PARAFAC receiver that estimates the \ac{UT}-\ac{MIS} channel, the \ac{MIS}-\ac{BS} channel, and the position-dependent movable-layer response through conditional least-squares updates. Thus, the receiver can operate without prior knowledge of the movable-layer phase matrix.

    \item We discuss the identifiability and computational complexity of the proposed receiver and evaluate how the number of training slots affects the channel estimation accuracy under the considered \ac{MIS} constraints.
\end{itemize}

\textit{Notation and Properties}: Vectors are denoted by boldface lowercase letters, e.g., $\ma{a}$, matrices by boldface uppercase letters, e.g., $\ma{A}$, and tensors by calligraphic letters, e.g., $\ten{A}$. The transpose and Moore-Penrose pseudo-inverse of $\ma{A}$ are denoted by $\ma{A}^T$ and $\ma{A}^{\dagger}$, respectively. 
The operator $\operatorname{diag}(\ma{a})$ forms a diagonal matrix from $\ma{a}$, whereas $\operatorname{Di}_i(\ma{A})$ denotes a diagonal matrix from the $i$-th row of $\ma{A}$. The Frobenius norm is represented by $\|\cdot\|_F$. The symbols $\otimes$ and $\diamond$ denote the Kronecker and Khatri-Rao products, respectively. The operator $\operatorname{vec}(\cdot)$ vectorizes a matrix, while $\operatorname{unvec}_{I\times J}(\cdot)$ reshapes a vector into an $I\times J$ matrix. The entries, rows, and columns of $\ma{A}$ are denoted by $A_{i,j}$, $\ma{A}_{i,:}$, and $\ma{A}_{:,j}$, respectively.



                        

\begin{figure}
    \centering
    \includegraphics[width=1\linewidth]{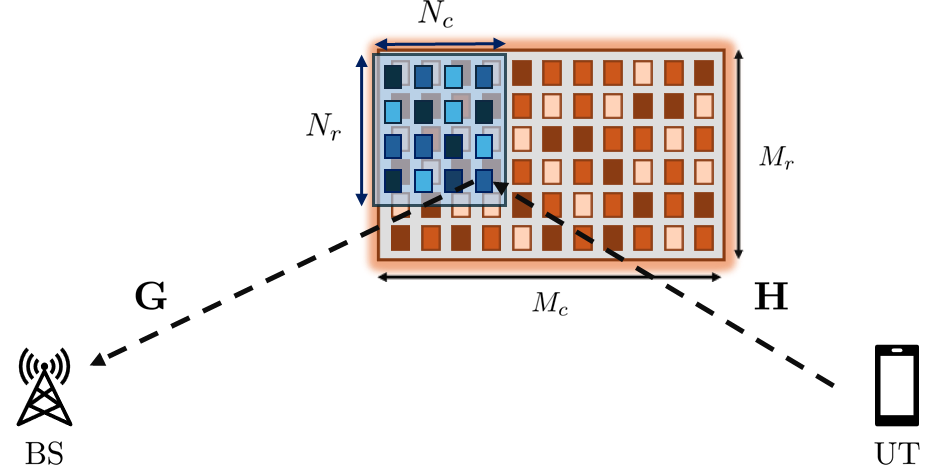}
    \caption{MIMO communications system assisted by MIS}
    \label{fig:system_model}
\end{figure}

\section{System and Signal Models}
We consider an uplink \ac{MIS}-assisted MIMO system in which a \ac{UT} with $N_t$ transmit antennas transmits pilot symbols to a \ac{BS} with $N_r$ receive antennas. The direct \ac{UT}-\ac{BS} link is assumed to be unavailable, so the transmission is assisted by the dual-layer \ac{MIS} illustrated in Figure~\ref{fig:system_model}. 
The \ac{MIS} architecture is based on a dual-layer transmissive configuration that synthesizes dynamic beam patterns via mechanical displacement rather than electronic tuning. It consists of a stationary reference grid, denoted by MS1, and a smaller movable layer, denoted by MS2, which slides over the surface of MS1 in discrete steps. The fixed layer, denoted by MS1, has $M$ transmissive elements, whereas the movable layer, denoted by MS2, has $N$ elements, with $N<M$\footnote{If $N=M$ and the coupling channel between MS1 and MS2 is explicitly considered, the architecture reduces to a two-layer stacked metasurface.}.
Specifically, MS1 is modeled as a uniform rectangular array with $M_r$ rows and $M_c$ columns, such that $M = M_r \times M_c$. The $m$-th element of MS1 belongs to the set $\mathcal{M} = \{1, \dots, M\}$ and is indexed as $m = (m_r - 1)M_c + m_c$, where $m_r \in \{1, \dots, M_r\}$ and $m_c \in \{1, \dots, M_c\}$. Similarly, MS2 has $N_r$ rows and $N_c$ columns, yielding $N = N_r \times N_c$ elements. The $n$-th element of MS2 belongs to the set $\mathcal{N} = \{1, \dots, N\}$ and is indexed as $n = (n_r - 1)N_c + n_c$, where $n_r \in \{1, \dots, N_r\}$ and $n_c \in \{1, \dots, N_c\}$. Unlike the approach in \cite{paperMIS}, which assumes that the phase-shift vectors $\ma{\phi} \in \mathbb{C}^{M \times 1}$ and $\ma{\theta} \in \mathbb{C}^{N \times 1}$ associated with MS1 and MS2, respectively, remain constant, we assume that $\ma{\phi}$ varies from block to block, while $\ma{\theta}$ remains fixed and only the position of the movable panel changes.

\begin{figure}[!t]
    \centering
    \includegraphics[width=1\linewidth]{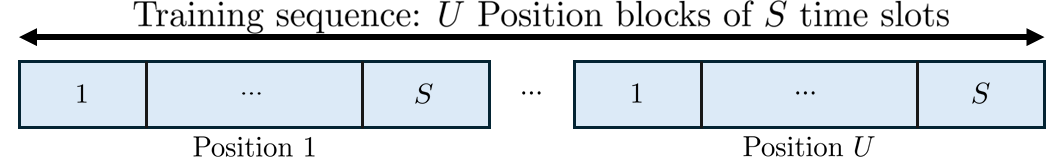}
    \caption{Training protocol}
    \label{fig:training_protocol}
\end{figure}

 During training, MS1 applies $S$ predefined phase-shift patterns, while MS2 explores $U$ discrete positions over the fixed grid, as illustrated in Figure~\ref{fig:training_protocol}.
\noindent
Since the \ac{MIS} generates different effective responses through differential position shifting, the received signal depends jointly on the time slot $s$ and on the mechanical state $u$. The received signal at the \ac{BS} during the $s$-th slot and $u$-th position is given by
\begin{equation}
    \mathbf{Y}[s,u] = \mathbf{G}\ma{\Phi}_m[u]\ma{\Phi}_f[s]\mathbf{H}\mathbf{X} + \mathbf{N}[s,u]\;,
\end{equation}
where the phase-shift matrices are defined as
\begin{align}
\mathbf{\Phi}_f[s] &= \text{diag}(\ma{\phi}_f[s]), \quad s = 1, \dots, S,
\label{eq:phi_f}\\
\mathbf{\Phi}_m[u] &= \text{diag}(\ma{\phi}_m[u]), \quad u = 1, \dots, U.
\label{eq:phi_m}
\end{align}
The matrices $\mathbf{H} \in \mathbb{C}^{M \times N_t}$ and $\mathbf{G} \in \mathbb{C}^{N_r \times M}$ denote the \ac{UT}-\ac{MIS} and \ac{MIS}-\ac{BS} channels, respectively. The pilot matrix is $\mathbf{X} = [\mathbf{x}_1 \; \mathbf{x}_2 \; \dots \; \mathbf{x}_T] \in \mathbb{C}^{N_t \times T}$, and $\mathbf{N}[s,u]$ represents additive white Gaussian noise. Assuming that the pilot matrix is orthogonal, the received signal can be filtered as
\begin{equation}
\label{receivedsignal1}
    \tilde{\mathbf{Y}}[s,u] \triangleq \mathbf{Y}[s,u]\mathbf{X}^{H} \approx \mathbf{G} \mathbf{\Phi}_m[u] \mathbf{\Phi}_f[s] \mathbf{H} + \tilde{\mathbf{N}}[s,u],
\end{equation}
\noindent where $\tilde{\mathbf{N}}[s,u] = \mathbf{N}[s,u]\mathbf{X}^{H}$ represents the filtered noise matrix. In this work, $\mathbf{\Phi}_f[s]$ is known because it is imposed by the fixed-layer training design, whereas $\mathbf{\Phi}_m[u]$ is assumed to be unknown at the receiver. This models the practical case in which the effective phase response of the movable layer is unavailable or only imperfectly calibrated. The noiseless version of \eqref{receivedsignal1} can be interpreted as the $(s,u)$-th matrix slice of a fourth-order tensor $\overline{\mathcal{Y}} \in \mathbb{C}^{N_r \times N_t \times S \times U}$, which follows a PARAFAC model given by
\noindent
\begin{equation}
\overline{\mathcal{Y}} = \mathcal{I}_{4,M} \times_1 \mathbf{G} \times_2 \mathbf{H} \times_3 \mathbf{\Phi}_f
\times_4 \mathbf{\Phi}_m\in \mathbb{C}^{N_r \times N_t \times S \times U}
\label{EQ:n-mode notation}
\end{equation}
\noindent
The cascaded channel is defined as $\mathbf{H}_c = (\mathbf{H}^T \diamond \mathbf{G})$.

\subsection{Differential Position Shifting}
\label{movimentoMIS}
Following the movable-surface architecture introduced in \cite{paperMIS}, we use the displacement of MS2 relative to the fixed MS1 grid as a controllable source of spatial diversity for channel estimation. Instead of electronically updating all reflecting elements, the proposed training protocol changes the relative position of MS2, so that each mechanical state produces a different effective phase profile to be used in \eqref{eq:phi_m}.
The set of admissible positions is denoted by $\mathcal{U}=\{1,\dots,U\}$. Since MS2 slides over MS1 on a rectangular grid, the number of vertical and horizontal shifts is determined by the array dimensions as $U = U_r \times U_c$, with $U_r = M_r - N_r + 1, \quad U_c = M_c - N_c + 1$,
where $U_r$ and $U_c$ are the numbers of feasible shifts along the vertical and horizontal dimensions, respectively. Each position $u \in \mathcal{U}$ is associated with the two-dimensional displacement index $(u_r,u_c)$ according to
$u = (u_r - 1) U_c + u_c$,
with $u_r \in \{1,\dots,U_r\}$ and $u_c \in \{1,\dots,U_c\}$. For each position, only a subset of MS1 is covered by MS2. Hence, the effective response seen by the incident wave depends on the overlap between the static coefficients of the two layers. This position-dependent response is the mechanism that links the design of the \ac{MIS} to the multiple observations collected during training.

\subsection{Selection Matrix Modeling}
\label{SelectionMatrix}
To incorporate the previous displacement mechanism into the signal model, we adopt the selection matrix representation from \cite{paperMIS}. For a given position $u$, the equivalent phase-shift vector $\ma{\bar{\theta}}_u \in \mathbb{C}^{M \times 1}$ can be expressed as
\begin{equation}
\ma{\bar{\theta}}_u = \mathbf{S}_u \ma{\theta} + \mathbf{e}_u,
\end{equation}
where $\mathbf{S}_u \in \{0,1\}^{M \times N}$ selects the entries of MS1 that are covered by MS2 at position $u$, while $\mathbf{e}_u \in \mathbb{C}^{M \times 1}$ completes the response over the non-overlapping elements. The entries of these quantities are defined as
\begin{equation}
[\mathbf{S}_u]_{m,n} = 
\begin{cases} 
1, & \parbox{5cm}{if the $n$-th element of MS2 overlaps the $m$-th element of MS1}, \\
0, & \text{otherwise},
\end{cases}
\end{equation}
\begin{equation}
[\mathbf{e}_u]_m = 
\begin{cases} 
0, & \parbox{5cm}{if the $m$-th element of MS1 is covered by MS2}, \\
1, & \text{otherwise}.
\end{cases}
\end{equation}

Thus, the displacement index $u$ is translated into a structured phase profile with dimensions compatible with the fixed grid. In our model, this profile defines the diagonal entries of $\mathbf{\Phi}_m[u]$ in \eqref{eq:phi_m}. Although this model explains the physical origin of the movable-layer response, the receiver does not require the entries of $\mathbf{\Phi}_m[u]$ to be known. Instead, the collection of these position-dependent profiles is represented by an unknown factor matrix to be recovered from the received tensor. As a result, the $U$ mechanical states of the \ac{MIS} generate the fourth dimension of the received-signal tensor, which is later exploited by the proposed PARAFAC-based receiver.
\label{sec:Proposed S}

\section{Proposed Receiver}
We build upon the design and the selection mechanism described in Section \ref{movimentoMIS} to formulate the proposed receiver. The different mechanical states of the \ac{MIS} provide multiple position-dependent observations, which are arranged together with the fixed-layer phase profiles into the fourth-order tensor introduced in \eqref{EQ:n-mode notation}. We then exploit the algebraic structure of the received pilot tensor to estimate the channel factors through a PARAFAC-based alternating least squares procedure \cite{comon_2009}.

Let $\mathbf{C}\in\mathbb{C}^{S\times M}$ collect the known fixed-layer phase profiles associated with $\mathbf{\Phi}_f[s]$, while $\mathbf{D}\in\mathbb{C}^{U\times M}$ collects the position-dependent phase profiles associated with $\mathbf{\Phi}_m[u]$. The matrix $\mathbf{D}$ is treated as unknown at the receiver, meaning that the movable-layer phase response is estimated rather than assumed to be available. 
The receiver estimates $\mathbf{G}$, $\mathbf{H}^T$, and $\mathbf{D}$ by solving the following multilinear problem:
\begin{equation}
(\hat{\mathbf{G}}, \hat{\mathbf{H}}^T, \hat{\mathbf{D}}) = \underset{\mathbf{G}, \mathbf{H}^T, \mathbf{D}}{\arg \min} \left\| \mathcal{Y} - [[\mathbf{G}, \mathbf{H}^T, \mathbf{C}, \mathbf{D}]] \right\|_F^2,
\label{eq:optimization_problem}
\end{equation}

The optimization in \eqref{eq:optimization_problem} is solved by exploiting the following $n$-mode unfoldings of the received tensor:
\noindent
\begin{equation}
[\mathcal{Y}]_{(1)} = \mathbf{G} (\mathbf{D} \diamond \mathbf{C} \diamond \mathbf{H}^T)^T
\label{eq:unfolding_1}
\end{equation}
\noindent
\begin{equation}
[\mathcal{Y}]_{(2)} = \mathbf{H}^T (\mathbf{D} \diamond \mathbf{C} \diamond \mathbf{G})^T
\label{eq:unfolding_2}
\end{equation}
\noindent
\begin{equation}
[\mathcal{Y}]_{(4)} = \mathbf{D} (\mathbf{C} \diamond \mathbf{H}^T \diamond \mathbf{G})^T
\label{eq:unfolding_4}
\end{equation}

Based on the unfoldings in \eqref{eq:unfolding_1}, \eqref{eq:unfolding_2}, and \eqref{eq:unfolding_4}, the estimates of $\hat{\mathbf{G}}$, $\hat{\mathbf{H}}^T$, and $\hat{\mathbf{D}}$ are obtained through the \ac{TALS} algorithm \cite{comon_2009}. At each iteration, two factors are fixed while the remaining one is updated by solving one of the following conditional LS subproblems:
\begin{eqnarray}
\hspace{-2ex}&&\hat{\mathbf{G}} = \underset{\mathbf{G}}{\arg\min} \quad \left\| [\mathcal{Y}]_{(1)} - \mathbf{G}(\mathbf{D} \diamond \mathbf{C} \diamond \mathbf{H}^T)^T \right\|_\text{F}^2,
\label{eq:ls_G}\\
\hspace{-2ex}&&\hat{\mathbf{H}}^T = \underset{\mathbf{H}^T}{\arg\min} \quad \left\| [\mathcal{Y}]_{(2)} - \mathbf{H}^T(\mathbf{D} \diamond \mathbf{C} \diamond \mathbf{G})^T \right\|_\text{F}^2,
\label{eq:ls_H}\\
\hspace{-2ex}&&\hat{\mathbf{D}} = \underset{\mathbf{D}}{\arg\min} \quad \left\| [\mathcal{Y}]_{(4)} - \mathbf{D}(\mathbf{C} \diamond \mathbf{H}^T \diamond \mathbf{G})^T \right\|_\text{F}^2,
\label{eq:ls_D}
\end{eqnarray}
the solutions of which are respectively given by:
\begin{equation}
\hat{\mathbf{G}} = [\mathcal{Y}]_{(1)} \left[ (\mathbf{D} \diamond \mathbf{C} \diamond \mathbf{H}^T)^T \right]^\dagger,
\label{eq:sol_G}
\end{equation}
\begin{equation}
\hat{\mathbf{H}}^T = [\mathcal{Y}]_{(2)} \left[ (\mathbf{D} \diamond \mathbf{C} \diamond \mathbf{G})^T \right]^\dagger,
\label{eq:sol_H}
\end{equation}
\begin{equation}
\hat{\mathbf{D}} = [\mathcal{Y}]_{(4)} \left[ (\mathbf{C} \diamond \mathbf{H}^T \diamond \mathbf{G})^T \right]^\dagger,
\label{eq:sol_D}
\end{equation}
\noindent

The resulting receiver follows the standard alternating structure commonly adopted in PARAFAC-based channel estimation methods: each factor matrix is updated from a single tensor unfolding, while the remaining factors are kept fixed. The update of $\mathbf{D}$ in \eqref{eq:sol_D} is the key step that enables operation under an unknown movable-layer phase response, since it extracts this response from the fourth-mode unfolding instead of relying on prior calibration. The procedure is summarized in Algorithm~\ref{algorithm1}.
\begin{algorithm}[!t]
\caption{TALS-PARAFAC receiver for MIS-assisted MIMO channel estimation}
\label{algorithm1}
\small
\textbf{Inputs:} received tensor $\mathcal{Y}$, known phase matrix $\mathbf{C}$, maximum number of iterations $I_{\max}$, and threshold $\epsilon$. The movable-layer phase matrix $\mathbf{D}$ is unknown.\;
Set $i=0$ and randomly initialize $\hat{\mathbf{H}}^T_{(0)}$ and $\hat{\mathbf{D}}_{(0)}$.\;
Set $\xi_{(0)}=\left\|\mathcal{Y}\right\|_F^2$.\;
\For{$i=1:I_{\max}$}{
    Update $\hat{\mathbf{G}}_{(i)}$ via LS as\;
    \begin{equation*}
    \hat{\mathbf{G}}_{(i)} = [\mathcal{Y}]_{(1)}\left[(\hat{\mathbf{D}}_{(i-1)}\diamond\mathbf{C}\diamond\hat{\mathbf{H}}^T_{(i-1)})^T\right]^\dagger.
    \end{equation*}
    Update $\hat{\mathbf{H}}^T_{(i)}$ via LS as\;
    \begin{equation*}
    \hat{\mathbf{H}}^T_{(i)} = [\mathcal{Y}]_{(2)}\left[(\hat{\mathbf{D}}_{(i-1)}\diamond\mathbf{C}\diamond\hat{\mathbf{G}}_{(i)})^T\right]^\dagger.
    \end{equation*}
    Update $\hat{\mathbf{D}}_{(i)}$ via LS as\;
    \begin{equation*}
    \hat{\mathbf{D}}_{(i)} = [\mathcal{Y}]_{(4)}\left[(\mathbf{C}\diamond\hat{\mathbf{H}}^T_{(i)}\diamond\hat{\mathbf{G}}_{(i)})^T\right]^\dagger.
    \end{equation*}
    Compute the reconstructed tensor\;
    \begin{equation*}
    \hat{\mathcal{Y}}_{(i)} = [[\hat{\mathbf{G}}_{(i)},\hat{\mathbf{H}}^T_{(i)},\mathbf{C},\hat{\mathbf{D}}_{(i)}]].
    \end{equation*}
    Compute the reconstruction error $\xi_{(i)}=\left\|\mathcal{Y}-\hat{\mathcal{Y}}_{(i)}\right\|_F^2$.\;
    \If{$i > 1$ \textbf{and} $|e(i) - e(i-1)| \leq \epsilon$}{
        \textbf{break}\;
    }
    
}
Return $\hat{\mathbf{G}}_{(i)}$, $\hat{\mathbf{H}}_{(i)}$, and $\hat{\mathbf{D}}_{(i)}$.\;
\end{algorithm}
The updates in \eqref{eq:sol_G}--\eqref{eq:sol_D} are repeated until convergence or until a maximum number of iterations is reached. The convergence metric is defined as the relative variation of the reconstruction error between two consecutive iterations, i.e., $\eta = |\xi_{(i)}-\xi_{(i-1)}|/\xi_{(i-1)}$, where $\xi_{(i)} = \left\| \mathcal{Y} - \hat{\mathcal{Y}}_{(i)} \right\|_F^2$. The algorithm stops when $\eta \leq \epsilon$, yielding the estimated channel factors and the position-dependent MIS response.

\textit{Identifiability and Computational Complexity}:
The conditional LS problems in \eqref{eq:ls_G}--\eqref{eq:ls_D} are well posed when the Khatri-Rao matrices associated with the updates of $\mathbf{G}$, $\mathbf{H}^T$, and $\mathbf{D}$, namely $(\mathbf{D}\diamond\mathbf{C}\diamond\mathbf{H}^T)$, $(\mathbf{D}\diamond\mathbf{C}\diamond\mathbf{G})$, and $(\mathbf{C}\diamond\mathbf{H}^T\diamond\mathbf{G})$, have full column rank. A necessary dimensional requirement for these matrices to be full column rank is
\begin{equation}
USN_t \geq M, \quad USN_r \geq M, \quad SN_tN_r \geq M.
\end{equation}
In practice, for rich scattering channels and assuming the phase profiles in $\mathbf{C}$ and $\mathbf{D}$ are sufficiently diverse, the Khatri-Rao products are full column rank with high probability whenever the above conditions are satisfied. 

The computational cost of the proposed receiver is dominated by the pseudoinverses in the LS updates. For a tall matrix $\mathbf{A}\in\mathbb{C}^{K\times M}$, a standard LS/SVD-based pseudoinverse has dominant cost $\mathcal{O}(KM^2+M^3)$, leading to a per-iteration complexity of $\mathcal{O}\left(M^2\left[US(N_t+N_r)+SN_tN_r\right]+3M^3\right)$ for the three TALS updates and a total complexity of $\mathcal{O}\left(I\left\{M^2\left[US(N_t+N_r)+SN_tN_r\right]+3M^3\right\}\right)$ after $I$ iterations. Therefore, the receiver complexity scales linearly with the number of TALS iterations and with the number of states $U$, while the identifiability of the LS subproblems is mainly controlled by the training length $S$, the number of positions $U$, and the conditioning of the designed phase profiles.

\section{Simulation Results}\label{results}

In this section, we evaluate the performance of the proposed PARAFAC-based channel estimation algorithm for the MIS-assisted system. The performance metric is the \ac{NMSE} of the estimated factors $\hat{\mathbf{H}}^T$, $\hat{\mathbf{G}}$, and $\hat{\mathbf{D}}$, as well as of the reconstructed cascaded channel. Unless otherwise stated, we consider $N_r = 8$, $N_t = 4$, $S \in \{32,48,64\}$, and $10^3$ Monte Carlo runs. The fixed and movable MIS layers are configured with $M = 64$ $(8\times8)$ and $N = 4$ $(2\times2)$ transmissive elements, respectively, providing a total of $U = 49$ possible mechanical positions.

\begin{figure}[!t]
    \centering
    \includegraphics[scale=0.49]{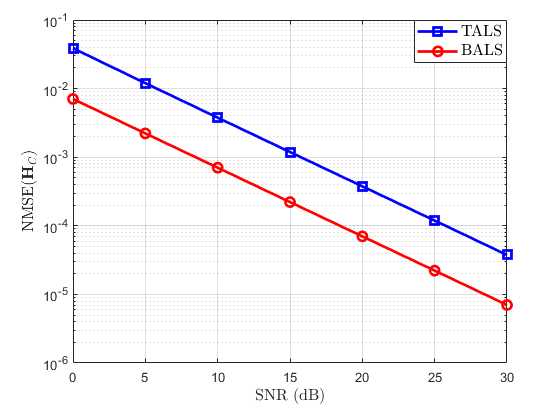} 
    \caption{NMSE of the reconstructed cascaded channel vs. SNR for the proposed \ac{TALS} receiver and the \ac{BALS} benchmark.}
    \label{fig:BALSxTALS}
\end{figure}

Figure \ref{fig:BALSxTALS} compares the NMSE of the reconstructed cascaded channel obtained with the proposed \ac{TALS} receiver and with the conventional \ac{BALS} benchmark. Both methods exhibit the expected monotonic reduction in NMSE as SNR increases, confirming that the tensor-based receiver successfully exploits the two-time-scale training scheme to extract accurate channel estimates. The \ac{BALS} benchmark achieves a lower NMSE over the entire SNR range. However, this benchmark assumes perfect knowledge of the movable-RIS training matrix at the receiver, whereas the proposed \ac{TALS} receiver treats it as unknown and jointly estimates it with the channel factors.
\begin{figure}[!t]
    \centering
    \includegraphics[scale=0.49]{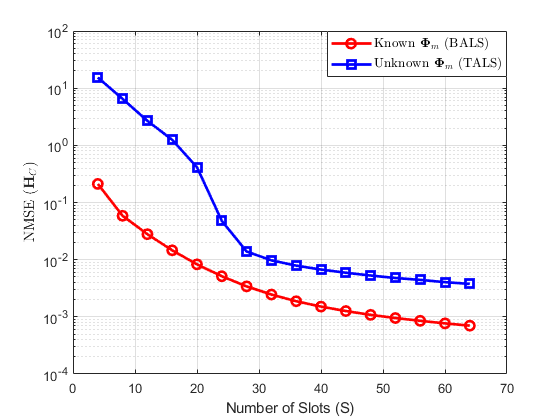}
    \caption{Impact of the number of training slots on the cascaded-channel reconstruction at SNR $=10$ dB.}
    \label{fig:SlotsComposto}
\end{figure}
In Fig. \ref{fig:SlotsComposto}, we study the impact of the number of training slots on the cascaded channel reconstruction at SNR $=10$ dB. The curve shows a pronounced transition as $S$ increases. Furthermore, while \ac{TALS} operates under a more challenging scenario where the mobile MIS phase shifts ($\mathbf{\Phi}_m$) are unknown, the traditional \ac{BALS} baseline exploits full knowledge of these phase profiles. Consequently, \ac{BALS} achieves a lower NMSE, highlighting the performance trade-off between assuming ideal knowledge of the channel phase and deploying a more flexible, blind estimation framework. For small values of $S$, the number of observations is insufficient to obtain a well-conditioned tensor factorization, resulting in a high NMSE. Once the training length is sufficiently large, the additional time diversity produces a sharp decrease in NMSE. For larger values of $S$, the curve decreases more slowly, indicating a regime in which the performance becomes mainly limited by noise and residual algorithmic errors rather than by the number of observations.

\begin{figure}[!t]
    \centering
    \includegraphics[scale=0.49]{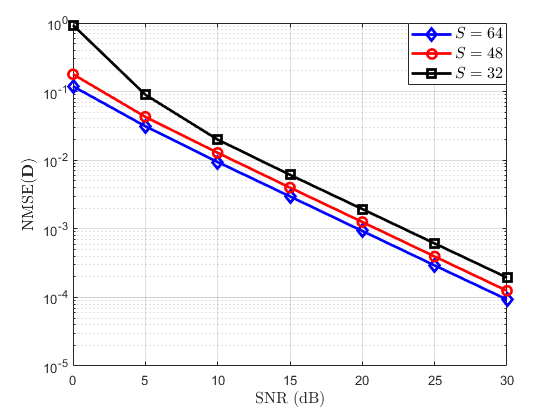}
    \caption{NMSE of the estimated movable-layer response $\mathbf{D}$ for different numbers of training slots.}
    \label{fig:SlotsRIS}
\end{figure}

\begin{figure}[!t]
    \centering
    \includegraphics[scale=0.49]{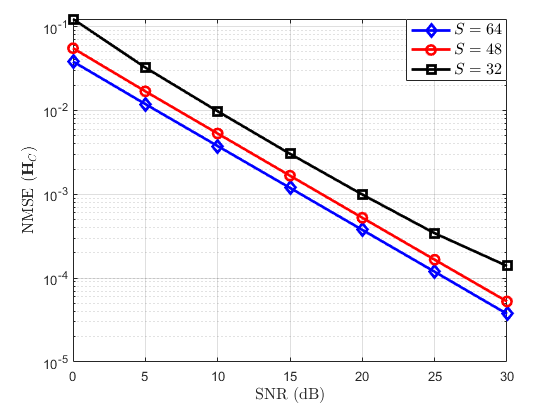}
    \caption{NMSE of the reconstructed cascaded channel for different numbers of training slots.}
    \label{fig:CanalComposto}
\end{figure}

Figures \ref{fig:SlotsRIS} and \ref{fig:CanalComposto} further illustrate the effect of the training length on the estimation of the movable-layer response $\mathbf{D}$ and on the reconstructed cascaded channel, respectively. It is worth emphasizing that $\mathbf{D}$ is not provided to the receiver in these simulations; it is recovered jointly with the channel matrices from the tensorized received signal. In both cases, increasing $S$ from $32$ to $64$ consistently improves the NMSE across the whole SNR range. This result is in line with the proposed system model, since a larger number of fixed-layer phase profiles increases the diversity of the received tensor along the training dimension and improves the identifiability of the factors. The nearly parallel behavior of the curves also indicates that the diversity gain from increasing $S$ is preserved across the range of SNR from low to high.

\noindent

\vspace{-0,2cm}
\section{Conclusion}
This paper proposed a new tensor-based channel estimation method for an uplink \ac{MIS}-assisted \ac{MIMO} system with fixed and movable transmissive metasurface layers. By jointly exploiting fixed-layer phase-shift patterns and movable-layer positions, the received pilots were modeled as a fourth-order PARAFAC tensor, from which a \ac{TALS} receiver was derived to estimate the \ac{UT}-\ac{MIS} channel, the \ac{MIS}-\ac{BS} channel, and the position-dependent movable-layer response. A key practical advantage of the proposed formulation is that the movable-layer phase matrix is not required at the receiver; instead, it is identified as a tensor factor from the received data. We verified that increasing the training length improves the NMSE of both the estimated movable-layer response and the reconstructed cascaded channel. Moreover, the comparison with the \ac{BALS} benchmark highlighted a trade-off between estimation accuracy and physical interpretability.

\renewcommand\baselinestretch{.84}

\bibliographystyle{IEEEtran}
\bibliography{IEEEexample}

\end{document}